\documentclass[11pt]{article} %
\usepackage{graphicx}
\usepackage{fancyhdr}
\usepackage{makeidx}
\usepackage{amssymb,amsmath}
\usepackage{enumerate}
\newcommand{\bd}{\begin{document}}
\newcommand{\ed}{\end{document}}
\newcommand{\bc}{\begin{center}}
\newcommand{\ec}{\end{center}}
\newcommand{\vs}{\vspace}
\newcommand{\hs}{\hspace}
\newcommand{\beq}{\begin{equation}}
\newcommand{\eeq}{\end{equation}}
\newcommand{\beqs}{\begin{eqn*}}
\newcommand{\eeqs}{\end{eqn*}}
\newcommand{\bq}{\begin{quote}}
\newcommand{\eq}{\end{quote}}
\newcommand{\lb}{\linebreak}
\newcommand{\mb}{\makebox}
\newcommand{\fb}{\framebox}
\newcommand{\mc}{\multicolumn}
\newcommand{\ben}{\begin{enumerate}}
\newcommand{\een}{\end{enumerate}}
\newcommand{\bit}{\begin{itemize}}
\newcommand{\eit}{\end{itemize}}
\newcommand{\ov}{\overline}
\newcommand{\un}{\underline}
\newcommand{\lt}{\left}
\newcommand{\rt}{\right}
\newcommand{\ba}{\begin{array}}
\newcommand{\ea}{\end{array}}
\newcommand{\beqa}{\begin{eqnarray}}
\newcommand{\eeqa}{\end{eqnarray}}
\newcommand{\beqas}{\begin{eqnarray*}}
\newcommand{\eeqas}{\end{eqnarray*}}
\newcommand{\bfg}{\begin{figure}}
\newcommand{\efg}{\end{figure}}
\newcommand{\pad}{\partial}
\newcommand{\nn}{\nonumber}
\newcommand{\la}{\leftarrow}
\newcommand{\ra}{\rightarrow}
\newcommand{\lgla}{\longleftarrow}
\newcommand{\lgra}{\longrightarrow}
\newcommand{\La}{\Leftarrow}
\newcommand{\Ra}{\Rightarrow}
\newcommand{\Lra}{\Leftrightarrow}
\newcommand{\Lgla}{\Longleftarrow}
\newcommand{\Lgra}{\Longrightarrow}
\renewcommand{\a}{\alpha}
\renewcommand{\b}{\beta}
\newcommand{\g}{\gamma}
\newcommand{\G}{\Gamma}
\renewcommand{\d}{\delta}
\newcommand{\D}{\Delta}
\newcommand{\e}{\epsilon}
\newcommand{\eps}{\epsilon}
\newcommand{\s}{\sigma}
\renewcommand{\l}{\lamda}
\newcommand{\m}{\mu}
\newcommand{\n}{\nu}
\renewcommand{\S}{\Sigma}
\newcommand{\p}{\pi}
\newcommand{\om}{\omega}
\newcommand{\Om}{\Omega}
\newcommand{\tri}{\triangle}
\newcommand{\ti}{\times}
\newcommand{\f}{\frac}
\newcommand{\ds}{\displaystyle}
\newcommand{\bm}[1]{\mb{{\boldmath $#1$}}}
\newcommand{\alter}[2]{\lt\{ \ba{ll}#1 \\ #2 \ea \rt.}
\newcommand{\alt}[4]{\lt\{ \ba{ll}#1 & \mb{if \, \,}#2 \\ #3 & \mb{}#4 \ea
    \rt.}
\newcommand{\altn}[4]{\lt\{ \ba{rl}#1 & \mb{if \, \,}#2 \\ #3 & \mb{}#4 \ea
    \rt.}
\newcommand{\altif}[4]{\lt\{ \ba{ll}#1 & \mb{if \, \,}#2 \\ #3 &
\mb{if \, \,}#4 \ea \rt.}
\newcommand{\altnif}[4]{\lt\{ \ba{rl}#1 & \mb{if \, \,}#2 \\ #3 &
\mb{if \, \,}#4 \ea \rt.}
\newcounter{algc}
\newcounter{romc}
\newcounter{Alphc}
\newcommand{\bl}{\begin{list}{{\it Step} ~\arabic{algc}~:} {\usecounter{algc}
                \setlength{\topsep}{0pt} \setlength{\itemsep}{0pt}}}
\newcommand{\el}{\end{list}}
\newcommand{\blr}{\begin{list}{~\roman{romc}~:} {\usecounter{romc}
                \setlength{\topsep}{0pt} \setlength{\itemsep}{0pt}}}
\newcommand{\elr}{\end{list}}
\newcommand{\bla}{\begin{list}{~\Alph{Alphc}~:} {\usecounter{Alphc}
                \setlength{\topsep}{0pt} \setlength{\itemsep}{0pt}}}
\newcommand{\ela}{\end{list}}

\newtheorem{theorem}{Theorem}
\begin{document}
\title{External Bias Dependent Direct To Indirect Bandgap Transition in Graphene Nanoribbon}
\author{Kausik Majumdar$^{1}$\thanks{Corresponding author, Email: kausik@ece.iisc.ernet.in},
K. V. R. M. Murali$^2$, Navakanta Bhat$^1$ and Yu-Ming Lin$^3$\\
$^1$ Department of Electrical Communication Engineering and \\Center
of Excellence in Nanoelectronics,
Indian Institute of Science, Bangalore-560012, India\\
$^2$ IBM Semiconductor Research and Development Center, Bangalore - 560045, India\\
$^3$ IBM T. J. Watson Research Center, Yorktown Heights, NY 10598, USA}
\date{}

\maketitle {\abstract In this work, using self-consistent
tight-binding calculations, for the first time, we show that a
direct to indirect bandgap transition is possible in an
semiconducting armchair graphene nanoribbon by the application of an
external bias along the width of the ribbon, opening up the
possibility of new device applications. With the help of Dirac
equation, we qualitatively explain this bandgap transition using the
asymmetry in the spatial distribution of the perturbation potential
produced inside the nanoribbon by the external bias. This is
followed by the verification of the bandgap trends with a numerical
technique using Magnus expansion of matrix exponentials. Finally, we
show that the carrier effective masses possess tunable sharp
characters in the vicinity of the bandgap transition points.}

\newpage
Graphene, the two dimensional allotrope of carbon, has drawn an
enormous amount of attention in the literature after its first
isolation on an oxide substrate \cite{ksn04,ksn05}. Apart from the
theoretical physics point of view \cite{akg07,ksn06}, graphene has
emerged as a possible candidate for different electronic devices
including field effect transistors \cite{mcl07}-\cite{yml10}.
However, the small bandgap of graphene reduces the controllability
of such devices and thus limits its widespread applications.
Graphene Nanoribbon (GNR), on the other hand, a quasi-one
dimensional strip of graphene, has been shown to provide a
significant bandgap \cite{kn96}-\cite{lb106} and hence is being
considered as a channel material in field effect transistors
\cite{bo06}-\cite{fmr08}.

More recently, it has been theoretically shown that the bandgap of a
graphene nanoribbon can be tuned significantly by the application of
an external field along the width of the ribbon
\cite{son06,dsn07,hr08} and at sufficiently large field, it is also
possible to collapse the gap of the ribbon. In this work, we extend
this result in a more generic way for both semiconducting and
metallic Armchair Graphene NanoRibbon (A-GNR) and demonstrate that
not only the magnitude of the bandgap, but the whole electronic
structure of the nanoribbon can be altered significantly depending
on the magnitude and polarity of the external bias. In particular,
we will show that it is possible to obtain a bias dependent direct
to indirect bandgap transition in such a nanoribbon. This kind of
control over the electronic structure by the application of an
external bias opens up the possibility of new device applications.

A schematic of the setup that we consider in this work is shown in
Fig. \ref{fig:schematic}, where an Armchair Graphene NanoRibbon
A-GNR of width $W$ is sandwiched between a left gate (G$_L$) and a
right gate (G$_R$). There is a third contact $V_c$ which keeps the
chemical potential of the A-GNR at zero. In this work, we will
primarily focus on semiconducting nanoribbons, i.e., the number of
dimers ($N$) along the width, is of the form $3M$ and $3M+1$. Note
that, $N=3M-1$ gives rise to metallic nanoribbons \cite{lb06,lb106}
and is briefly analyzed in this work. The gate terminal in each gate
stack is separated from the GNR by a dielectric. We assume the
Equivalent Oxide Thickness (EOT) of the gate dielectric to be 1nm.
The interfaces between the GNR and the gate dielectric are assumed
to be perfect, hence the electronic structure of the GNR is not
altered significantly. The gate dielectric confines the electrons
and holes in the GNR along the $x$ direction, however the $y$
component of the states can be obtained using longitudinal wave
vector $k_y\equiv k$. The work function of the gate metal is assumed
to be such that a zero flatband voltage is obtained. The external
biases $V_l$ and $V_r$ at the left and the right gates respectively
can be varied independently. The application of an external bias
changes the potential energy $U(x)=-q\phi(x)$ inside the GNR,
altering the electronic structure.

We now provide the details of the calculation procedure used to
investigate the effect of external bias on an A-GNR. The
self-consistent electronic structure of the A-GNR is determined by
using tight binding method (\cite{rs98, jcs54}) coupled with the
Poisson equation. Taking the left edge of the GNR at $x=0$ and the
plane of the GNR as $z=0$, the charge density is given by
$\rho(x,z)=qn(x,z)$, where $q$ is the electronic charge and $n(x,z)$
is obtained as the difference between the hole [$n_h(x,z)$] and
electron [$n_e(x,z)$] density as \beq\label{eq:n}
n(x,z)=2\left[\sum_{i,\bar{k}}(1-f(E_i(\bar{k})))|\psi_i^{\bar{k}}(x,z)|^2
-\sum_{j,\bar{k}}f(E_j(\bar{k}))|\psi_j^{\bar{k}}(x,z)|^2\right]
\eeq where $f(E)=\frac{1}{1+e^{(E-\mu)/k_BT}}$ is the Fermi-Dirac
probability at temperature $T$. Here $\bar{k}$ goes over the whole
first Brillouin Zone, $i$ and $j$ are the valence and conduction
band indices respectively. The chemical potential $\mu$, set by the
contact $V_c$, is taken to zero. $E_{i}(\bar{k})$ is the energy
eigenvalue of the state $(i,\bar{k})$ obtained from the tight
binding bandstructure taking only $p_z$ orbital into account, with
an intra-layer overlap integral, $S$ = 0.129 between two nearest
carbon atoms and the intra-layer hopping $t$ as $-$3.033eV
\cite{rs98}. Note that, the results obtained from the nearest
neighbor calculation are in close agreement with simulation that
take into account coupling terms up to the third nearest neighbor
(see supporting information). To obtain the wavefunction
$\psi_i^{\bar{k}}(x,z)$, we assume normalized Gaussian orbital as
the basis function, where the parameter of the basis function is
fitted using the parameter $S$. The wavefunctions are set to zero at
the dielectric interfaces indicating an infinite potential barrier.
Once self-consistency is achieved between the bandstructure
calculation and the Poisson equation for a given gate bias, the
energy eigenvalues at different $k$ points correspond to the
electronic structure of the A-GNR.

We take an A-GNR with $N=36$ ($W=4.55$nm) and consider three
representative bias conditions, namely, (i) $V_l=-V_r$, (ii) $V_l >
0$, $V_r=0$ and (iii) $V_l < 0$, $V_r=0$. We now present the results
in these three cases as shown in Fig.
\ref{fig:gap_eh}-\ref{fig:gap_vlr}.\\

\textbf{Case (i):} In this case, the two gate voltages are
anti-symmetric in nature, i.e., $V_l=-V_r=V_g$. We observe a
significant reduction of bandgap in Fig. \ref{fig:gap_eh}(a) and (b)
with an increase in $V_g$, and this has also been predicted in
\cite{dsn07,hr08}. It is observed that at sufficiently large $V_g$,
both the conduction and valence band edges shift from $k=0$, giving
a {\it `Mexican Hat'} shape around $k=0$. Fig. \ref{fig:gap_eh}(b)
clearly shows a threshold like behavior of the bandgap change
\cite{dsn07}, and as the band edges shift from $k=0$ (non-zero
$\Delta k$), the bandgap starts decreasing significantly with bias.
However, the particle-hole symmetry is almost conserved (the small
asymmetry in Fig. \ref{fig:gap_eh}(a) is due to the non-zero overlap
$S$ assumed between two nearest neighbor carbon atoms in the
honeycomb lattice) and hence the bandgap continues to remain direct
in nature, at any bias condition. Note that, the anti-symmetric bias
condition forces the GNR to retain its charge neutral condition with
similar electron and hole density, keeping the total effective
charge density very low. $\phi(x)$, dictated by the Poisson's
equation, thus remains almost linear (uniform field) along $x$, as
shown in Fig. \ref{fig:gap_eh}(c). Note that bias dependent bandgaps
match very well with one of the previously published reports based
on non-selfconsistent calculations \cite{hr08} and this linearity of
$\phi(x)$ along the width of the nanoribbon is the reason of this
unexpected close match.

\textbf{Case (ii):} In this case, the left gate is kept at positive
bias, keeping the right gate grounded and the results are shown in
Fig. \ref{fig:gap_e}(a)-(c). The bandstructure shows a dramatic
change as compared with case (i). With an increase in $V_l=V_g$, the
conduction band minimum shifts from $k=0$, giving rise to a {\it
`Mexican Hat'} shape around $k=0$. However, the valence band maximum
continues to remain at $k=0$, irrespective of $V_g$. Thus, at any
$V_g$ for which $\Delta k > 0$, the GNR has an indirect bandgap. The
direct and indirect bandgap regions are indicated in Fig.
\ref{fig:gap_e}(b). The magnitude of the bandgap continues to show a
threshold-like behavior as before, with an increased sensitivity of
bandgap when the system becomes indirect. Note that, the spatial
distribution of $\phi(x)$ along the width of the nanoribbon is
severely non-linear (non-uniform field) to support the increased
electron density inside the GNR that arises due to the conduction
band edge moving closer to the chemical potential. We will later
point out that it is this strong non-linearity of $\phi(x)$ that
causes such a direct to indirect bandgap transition.

\textbf{Case (iii):} A similar case like (ii) can be constructed
where the left gate is at negative bias, with the right gate
grounded. The bandstructure in such a scenario is shown in Fig.
\ref{fig:gap_vlr}(a) where the conduction band minimum continues to
remain at $k=0$, and the valence band maximum shifts away from
$k=0$, depending on the external bias. In this case, the valence
band edge moves closer to the chemical potential resulting in a
relative increase in the hole density.

Note that the corrections due to the second and the third nearest
neighbor interactions contribute at relatively large values of $k$,
away from the zone center \cite{sr02}. However, the band edge shift
($\Delta k$) from the zone center ($k=0$), in all the above
scenarios, are small ($\sim 5\%$) compared to the size of the
Brillouin zone. Hence, the calculations with nearest neighbor
interactions are accurate enough to predict such direct to indirect
bandgap transition.

In Fig. \ref{fig:gap_vlr}(b), we generalize this result and show the
transition from direct to indirect bandgap in the ($V_l$,$V_r$)
space. We compute the absolute difference of the $k$ values of the
conduction band minimum and the valence band maximum for any
arbitrary combination of $V_l$ and $V_r$. This is plotted as a
function of ($V_l$,$V_r$) in Fig. \ref{fig:gap_vlr}(b). A zero value
(dark color) indicates direct bandgap, whereas a non-zero value
(lighter color) represents indirect bandgap region. We clearly
observe that in the ($V_l$,$V_r$) space, there are symmetric pockets
of indirect bandgap regions, with the chosen cases (ii and iii) are
the most favorable conditions to obtain such a bandgap transition.\\

In the case of a metallic A-GNR, it is interesting to note that an
asymmetric external electric field along the width opens a small
bandgap at the zone center. Fig. \ref{fig:metallic}(a) shows a
direct bandgap of $\sim 18$meV for a metallic A-GNR with $N=35$
under a bias of 2.8V at the left gate, while grounding the other.
However, as shown in Fig. \ref{fig:metallic}(b), at larger bias,
this bandgap tends to become indirect accompanied with a reduction
in its magnitude. We do not observe such an effect in the case of
anti-symmetric bias condition, in agreement with \cite{dsn07}.

The external bias dependent direct to indirect bandgap transition,
coupled with the change in magnitude of the bandgap can have
significant effects in phenomena including band-to-band tunneling,
electron-phonon interaction and optical properties. Such an external
bias dependent tailoring of the electronic structure can provide us
with the possibility of a wide variety of fascinating electronic and
optoelectronic device applications.\\

Now, to get more insights, we present a theoretical analysis of the
phenomenon by starting from the Dirac equation
\cite{lb06,lb106,dsn07}. We write the low energy states
$\Psi(\bar{r}) = e^{ik_0x}\psi_{+}(\bar{r}) +
e^{-ik_0x}\psi_{-}(\bar{r})$ in terms of smoothly varying envelop
$\psi=\{\psi_{+},\psi_{-}\}$. $\psi_{+}$ and $\psi_{-}$ have
components on the $A$ and $B$ sublattices in the honeycomb lattice
with $k_0=-4\pi/3a_0$ and $a_0=2.44$nm \cite{dsn07}. By making the
replacement $k_x\rightarrow -i\partial_x$ in the Dirac Hamiltonian
\cite{lb106}, we can write $H\psi=E\psi$ where the Hamiltonian ($H$) for
the nanoribbon is given as
\beq\label{eq:H}
H = \left( \begin{array}{cc}
H_{+} & 0  \\
0 & H_{-}
\end{array} \right)
\eeq
with $H_{\pm}=\pm i\hbar v\sigma_x\partial_x - \hbar vk\sigma_y
 -q\phi(x)\mathbf{I}$. Here, $\sigma$ are the Pauli matrices and
$v\approx10^6$m/s. To keep the analysis simple, we assume the
intra-layer coupling parameter $S$ to be zero. The armchair boundary
condition with ideal edges forces
\beq \label{eq:bc1} \psi_{+}(0) +
\psi_{-}(0)=0
\eeq
and
\beq\label{eq:bc2} \psi_{+}(W) +
e^{ik_0W}\psi_{-}(W)=0
\eeq
Now, we give a simple argument to show
why we observe a direct to indirect bandgap transition in setup (ii)
and (iii), whereas setup (i) provides direct bandgap independent of
external bias. If we write the full Hamiltonian $H$ by
discretization of space along $x$, we find,
\beq
Tr(H)=-4q\sum_j\phi(x_j)
\eeq
which is equal to zero in case (i) and
this holds good for any $k$. This is due to the anti-symmetric
nature of the external bias and hence of $\phi(x)$ about the mid
point of the nanoribbon. Now, using the fact that the sum of the
eigenvalues equals the trace of $H$, this condition forces the sum
of the energy eigenvalues at any $k$ to be zero. Thus the conduction
band and valence band remain symmetric about $\mu$, forcing the bandgap to be
direct at any external gate bias. However, in cases (ii) and (iii),
the asymmetric gate biases introduce consequent asymmetry in the
spatial distribution of $\phi(x)$ and hence force $Tr(H)$ to become
nonzero, allowing asymmetry in the conduction and the valence bands.
This manifests as a bandgap transition in the nanoribbon.

We now provide an independent numerical method derived from Eq. \ref{eq:H}
to re-calculate the bias dependent electronic structure and verify the trend
of direct to indirect bandgap transition obtained from tight binding calculations.
To do this, We rewrite $H\psi=E\psi$ as \cite{dsn07}
\beq
\partial_x \psi_{\pm}= \pm \zeta \psi_{\pm}
\eeq
where
\beq \zeta(x) = k\sigma_z - i\sigma_x(q\phi(x)+E)/\hbar v
\eeq
Since, in general, $\zeta(x)$ does not commute for two
different $x$, we can write the solutions in terms of Magnus series
\cite{mm54,sb09}:
\beq\label{eq:psi_m}
\psi_{\pm}(W)=e^{\theta_{\pm}}\psi_{\pm}(0)
\eeq
where $\theta_{\pm}=\sum_{j=1}^\infty(\pm1)^j\theta_j$. $\theta_{j}$ is
the $j^{th}$ term in the Magnus series with the first three terms
are given as \setlength{\arraycolsep}{-2.2em}
\begin{eqnarray}
\theta_1 = \int_0^W\zeta(x_1)dx_1, \nonumber\\
&&\theta_2 = \int_0^W dx_1\int_0^{x_1} dx_2[\zeta(x_1),\zeta(x_2)],\nonumber\\
&&\theta_3 = \int_0^W dx_1\int_0^{x_1} dx_2\int_0^{x_2} dx_3
([\zeta(x_1),[\zeta(x_2),\zeta(x_3)]]
+[\zeta(x_3),[\zeta(x_2),\zeta(x_1]])
\end{eqnarray}
\setlength{\arraycolsep}{5pt} Using Eqs. \ref{eq:bc1}, \ref{eq:bc2}
and \ref{eq:psi_m}, we obtain \beq (e^{\theta_{+}} -
e^{ik_0W}e^{\theta_{-}})\psi_{+}(0)=0\eeq To get non-trivial
solutions for $\psi_{+}(0)$, we obtain \beq\label{eq:det}
det\left[e^{\theta_{+}} - e^{ik_0W}e^{\theta_{-}}\right]=0 \eeq For
a given $k$, the set of values of $E$ satisfying Eq. \ref{eq:det}
gives the required energy eigenvalues, which can be found
numerically. We have verified that the results obtained using this
method show a direct to indirect bandgap transition in setup (ii)
and (iii) whereas the GNR continues to remain a direct bandgap
semiconductor in case (i).

As a special case, $\zeta(x)$ commutes for two different $x$ for
$k=0$ and hence $\theta_j(k$=$0)$ becomes zero for $j>1$. Hence, we
can readily observe from Eq. \ref{eq:det} that as long as $\int_0^W
\phi(x)dx = 0$, we do not have any change in the energy eigenvalues
at $k=0$ for any arbitrary $\phi(x)$. This is why we should not
expect any change in $E(k$=$0)$ for any external bias as long as
$V_l=-V_r$. Note that, in reality, as shown in Fig.
\ref{fig:gap_eh}(a), we do see a small change in $E(k$=$0)$ under
gate bias, which arises from non-zero overlap parameter $S$.
However, in case (ii) and (iii), nonzero $\int_0^W \phi(x)dx$
introduces a bias dependent upward or downward shift in the $E(k=0)$
value depending on the polarity of the terminal bias.

As a final comment, we extract the effective mass values at the
conduction band minimum and the valence band maximum for case (i)
and (ii) using the $E-k$ relationship obtained from self-consistent
tight binding calculations. The results are shown in Fig.
\ref{fig:mstar}(a)-(b). In both the cases, we observe strong
non-monotonic behavior of the effective mass values, both for the
electrons and the holes. In case (i) [Fig. \ref{fig:mstar}(a)], the
effective mass of the electrons follows that of the holes for both
small and large biases. However, at some intermediate gate bias,
where the band edges start shifting from $k=0$, we notice a
significant difference in the electron and the hole effective mass
values, though both of them strongly peak about that point. A
similar behavior is observed in the electron effective mass in case
(ii) [Fig. \ref{fig:mstar}(b)], which sharply peaks around the
direct to indirect band transition point indicating a ``flattening"
of the conduction band edge when it moves away from $k=0$. However,
we do not observe any such sharp peak in the hole effective mass
around the bandgap transition point. This sharp notch like behavior
of the effective masses (with almost an order of magnitude change)
is a unique feature of the influence of the external field on the
electronic structure of A-GNR where one can selectively ``slow down"
the carriers by choosing the appropriate external bias condition.

To conclude, using a self-consistent tight binding calculation, we
have demonstrated that it is possible to change the bandgap of an
semiconducting A-GNR from direct to indirect by adjusting the
external biases at the left and the right gate, opening up the
possibility of new device applications. Such a direct to indirect
bandgap transition has been explained, both qualitatively as well as
quantitatively, by starting from Dirac equation, to support the
findings obtained from tight-binding calculations. Finally, the
external bias dependent carrier effective masses have been shown to
have non-monotonic sharp behavior around the direct to indirect
bandgap transition point.

\textbf{Acknowledgement:} K. Majumdar and N. Bhat would like to
thank the Ministry of Communication and Information Technology,
Government of India, and the Department of Science and Technology,
Government of India, for their support.

\newpage
\bfg[htbp!] \bc
\includegraphics[scale=0.65]{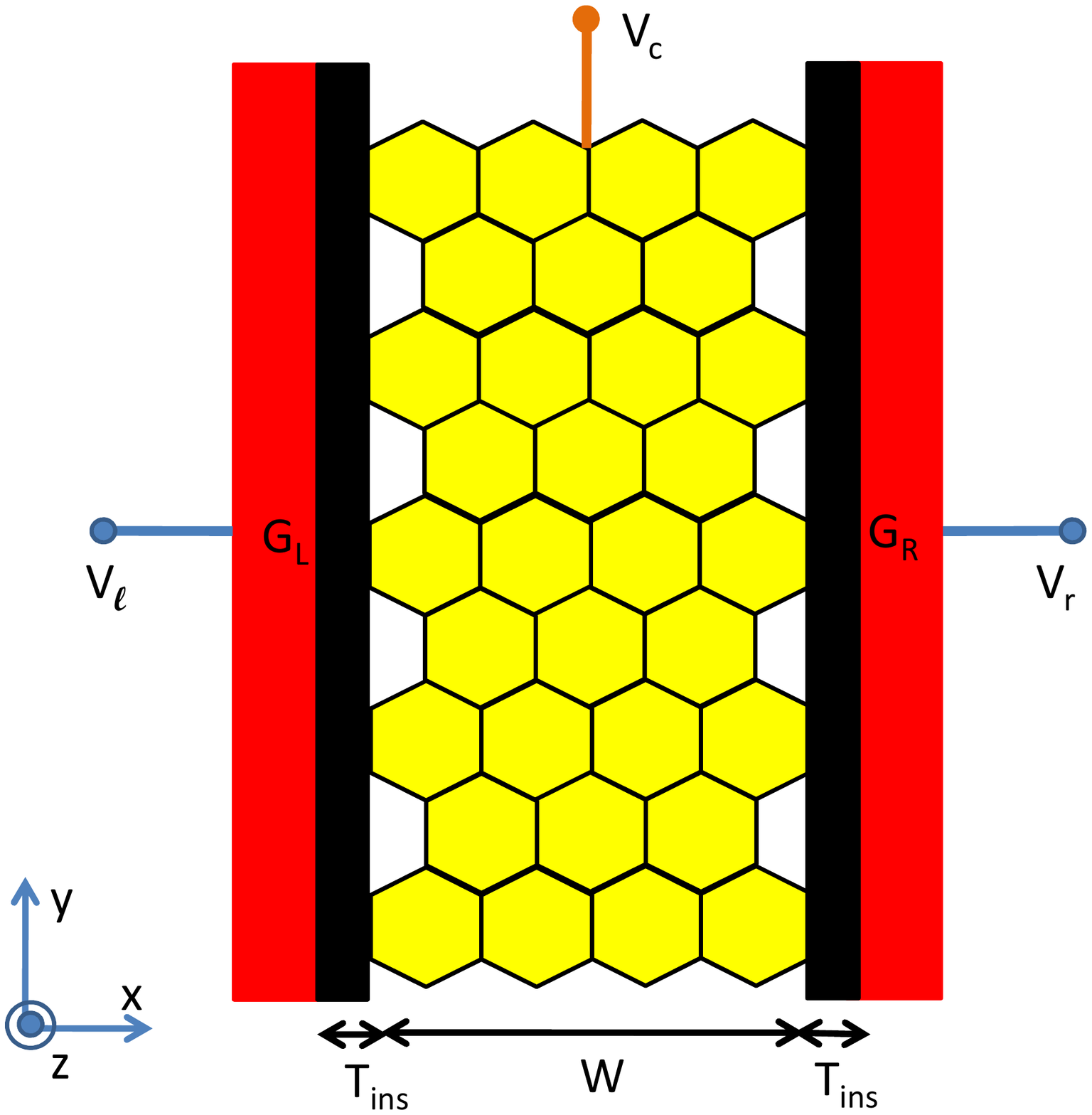}
\caption{A schematic of the A-GNR device with the external bias
along the width of the GNR. The chemical potential of the GNR is set
through the contact $V_c$ which is kept at zero bias (the reference
potential). The gate biases $V_l$ and $V_r$ create an external field
along $x$ inside the GNR which tunes the electronic structure of the
GNR.}\label{fig:schematic} \ec \efg

\newpage
\bfg[htbp!] \bc
\includegraphics[scale=0.65]{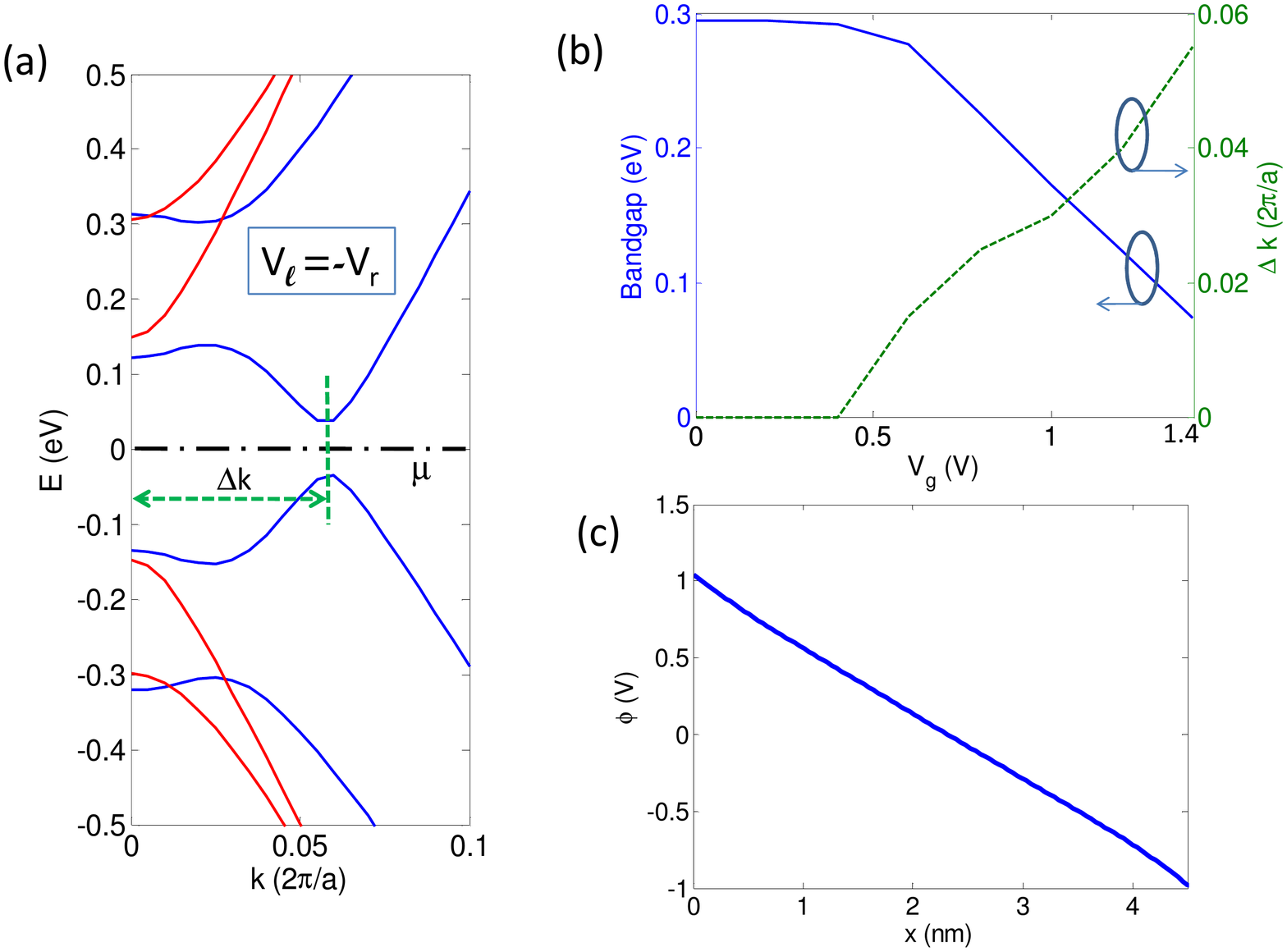}
\caption{The bandstructure of the A-GNR device ($N=36$) using
tight-binding method with the external biases at
$V_l$=$-V_r$=$V_g$=$1.4$V. The Chemical potential, $\mu$, is set to
zero in all the figures. (a): $E$-$k$ relationship (in blue) in the
vicinity of $k$=0 shows a significant reduction of bandgap, but the
gap remains direct. The red lines show the case when $V_g$=0. $k$ is
in terms of $\frac{2\pi}{a}$ where $a=3\alpha_0$ with
$\alpha_0=1.42${\AA} (b): Change in bandgap $E_g$ as a function of
$V_g$ shows a clear threshold-like behavior. In the same plot, we
show the shift ($\Delta k$) of the band edges from $k=0$. (c):
$\phi(x)$ plotted along $x$ inside the A-GNR shows an almost linear
spatial variation, indicating a uniform electric field along the
width of the ribbon.}\label{fig:gap_eh} \ec \efg

\newpage
\bfg[htbp!] \bc
\includegraphics[scale=0.65]{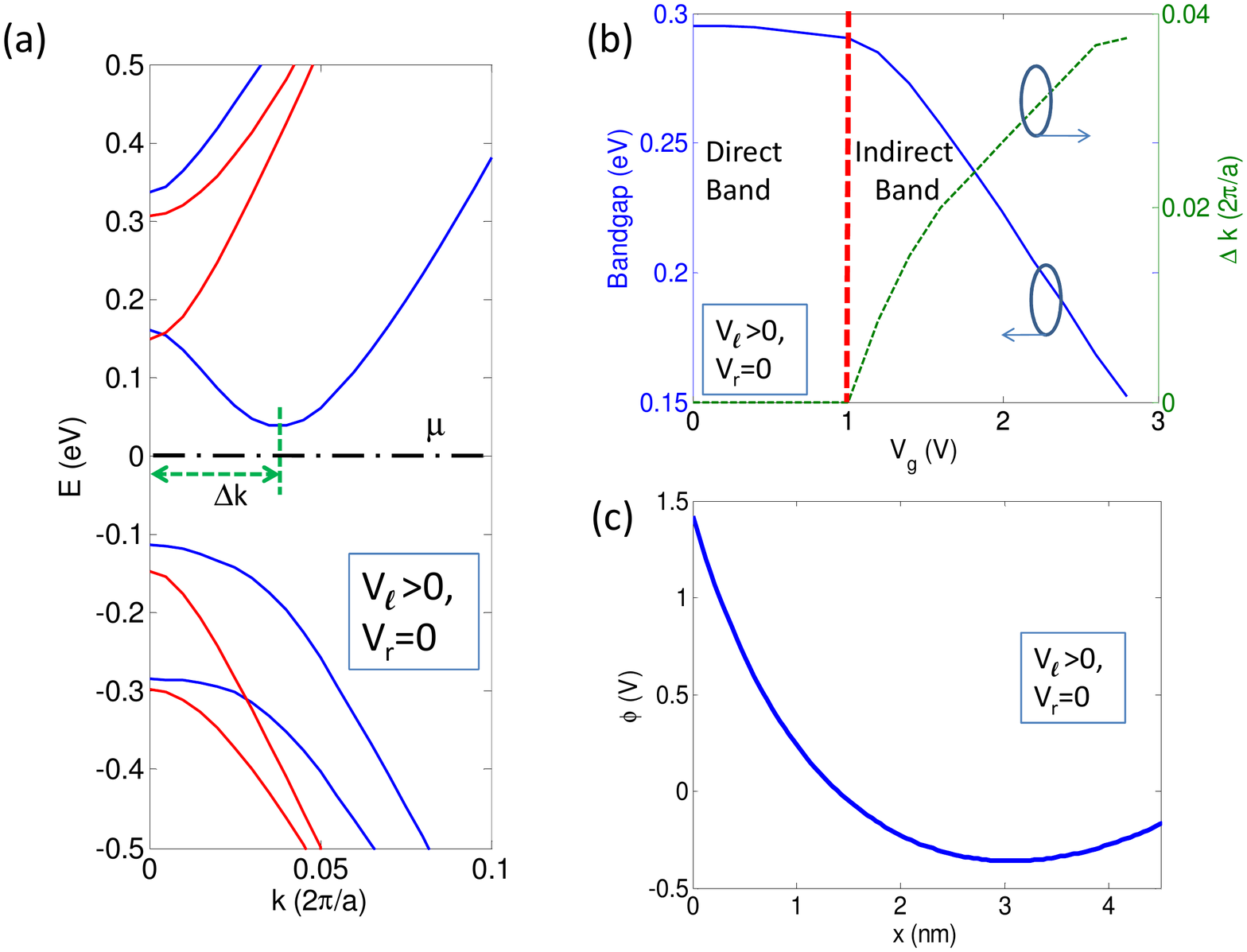}
\caption{The tight binding results of the A-GNR device with external
bias conditions, $V_l$=$V_g$=2.8V, $V_r$=0V. (a): $E$-$k$
relationship (in blue) in the vicinity of $k$=0 shows that the
conduction band minimum is shifted from $k$=0, while the valence
band maximum is at $k$=0, causing an indirect bandgap. The red lines
show the case when $V_g$=0. (b): Change in the bandgap $E_g$ as a
function of $V_g$ again shows a threshold-like behavior. In the same
plot, we show the shift ($\Delta k$) of the conduction band minimum
from $k=0$. (c): $\phi(x)$ plotted along $x$ in the A-GNR shows
significant nonlinearity in it spatial distribution, indicating a
non-uniform electric field. The Chemical potential $\mu$ is set to
zero in all the figures (a)-(c). }\label{fig:gap_e} \ec \efg

\newpage
\bfg[htbp!] \bc
\includegraphics[scale=0.65]{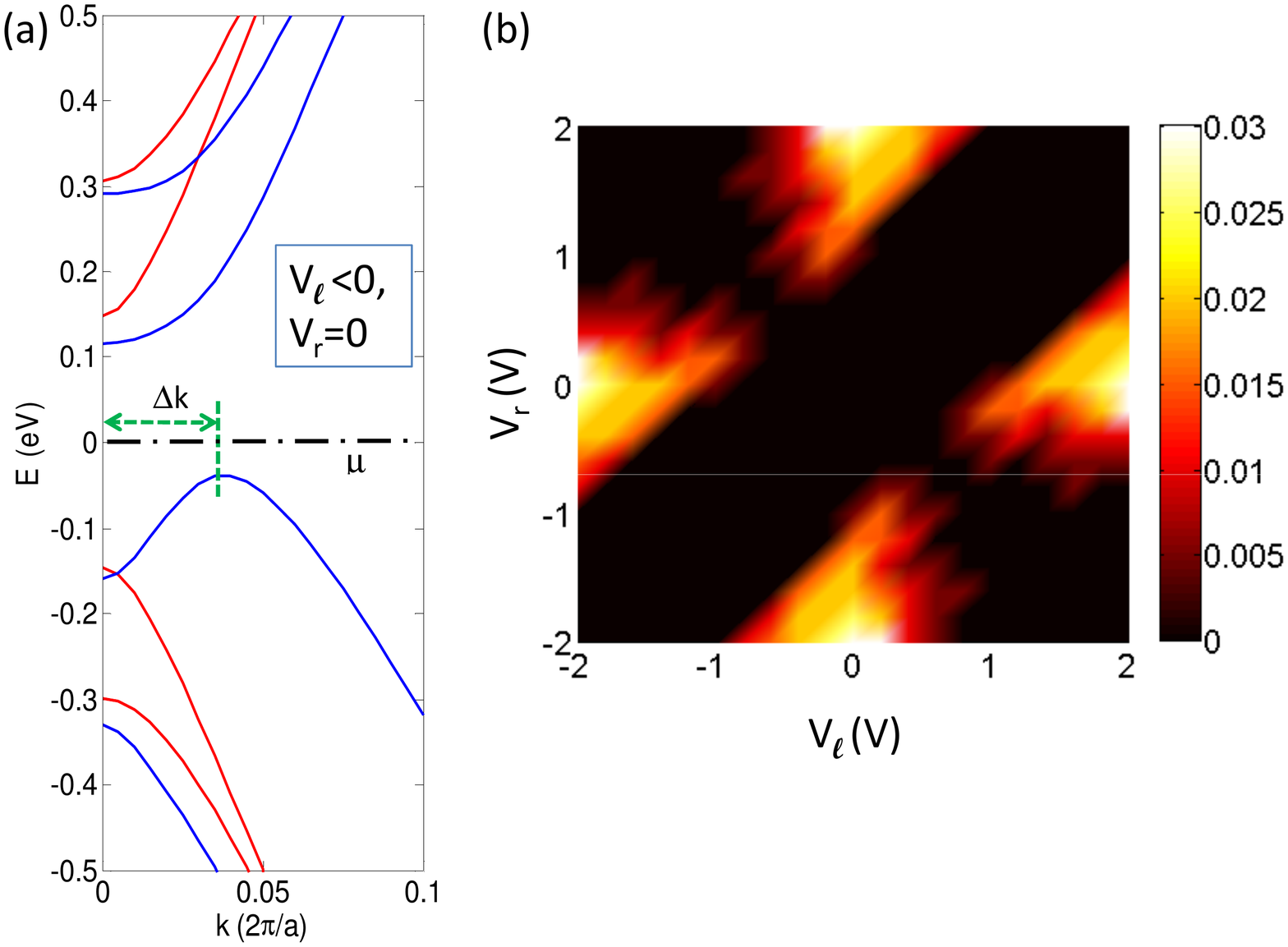}
\caption{(a): $E$-$k$ relationship (in blue) in the vicinity of
$k$=0 for $V_l$=$V_g$=$-2.8$V, $V_r$=0V shows that the conduction
band minimum remains at $k$=0, though the valence band maximum
shifted from $k$=0, causing an indirect bandgap. The red lines show
the electronic bands for the unbiased condition. (b): Plot of the
absolute of the difference between the $k$ values (in units of
$2\pi/a$) of the conduction band minimum and the valence band
maximum in the ($V_l$,$V_r$) space. A zero value of the same
indicates direct bandgap (dark color), whereas non-zero values
represent indirect bandgap (lighter color). We observe symmetric
pockets of indirect bandgap regions in the ($V_l$,$V_r$)
space.}\label{fig:gap_vlr} \ec \efg

\newpage
\bfg[htbp!] \bc
\includegraphics[scale=0.65]{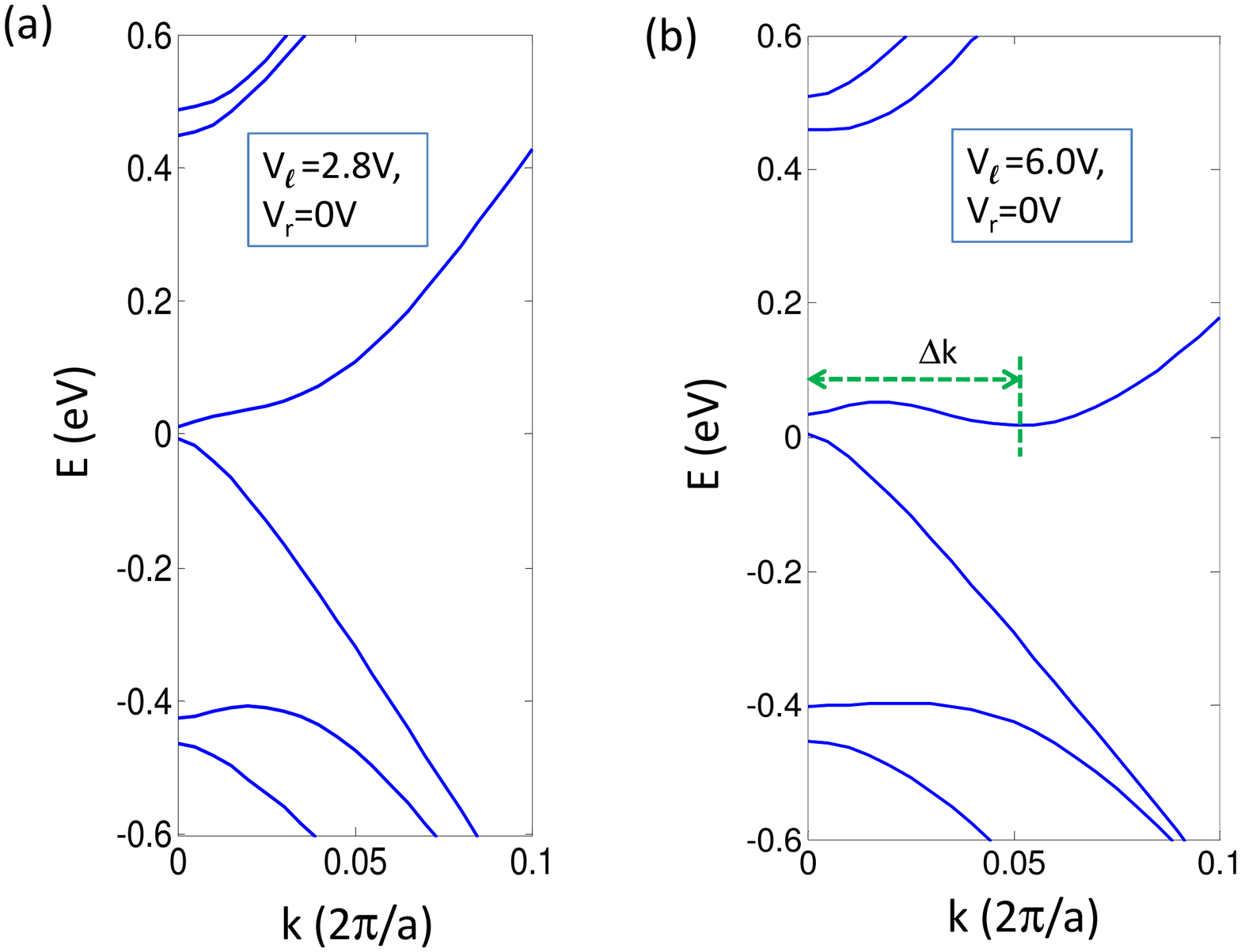}
\caption{(a): $E$-$k$ relationship in the vicinity of $k$=0 (for
$V_l$=$V_g$=$2.8$V, $V_r$=0V) shows the opening of a small direct
bandgap ($\sim$18meV) at the zone center, for a metallic A-GNR with
$N=35$. (b): The bandgap becomes indirect with a slight reduction in
its magnitude ($E_g\sim 14$meV) at $V_l=6$V.}\label{fig:metallic}
\ec \efg

\newpage
\bfg[htbp!] \bc
\includegraphics[scale=0.65]{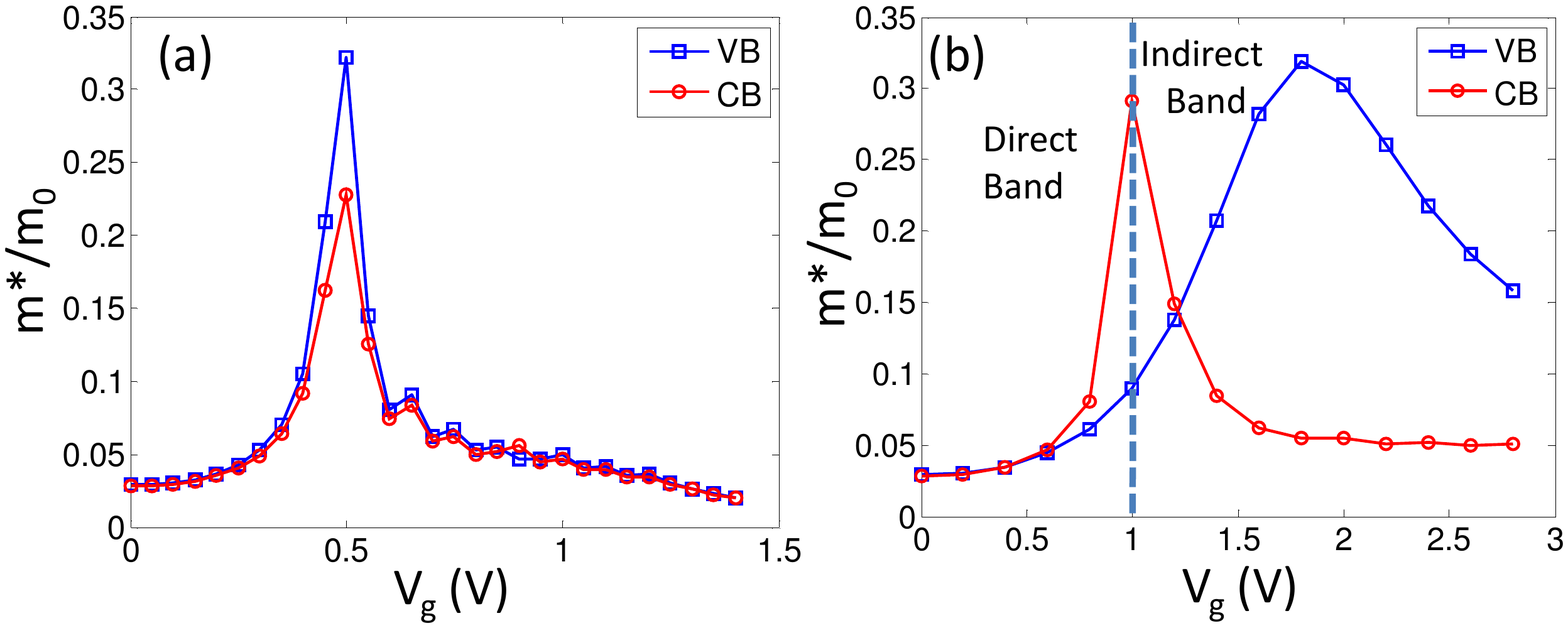}
\vs{-2in} \caption{(a): Effective mass at the conduction band
minimum and the valence band maximum as a function of external bias
in case (i) with $V_l=-V_r=V_g$. The effective masses are strongly
non-monotonic with the peak occurs in the vicinity of the bias point
where the band edges shift away from $k=0$. (b) Effective mass at
the conduction band minimum in case (ii) ($V_l=V_g>0, V_r=0$) shows
a sharp peak around direct to indirect bandgap transition
point.}\label{fig:mstar} \ec \efg

\newpage
\begin{figure}[hbt!]
\bc
\includegraphics[scale=0.6]{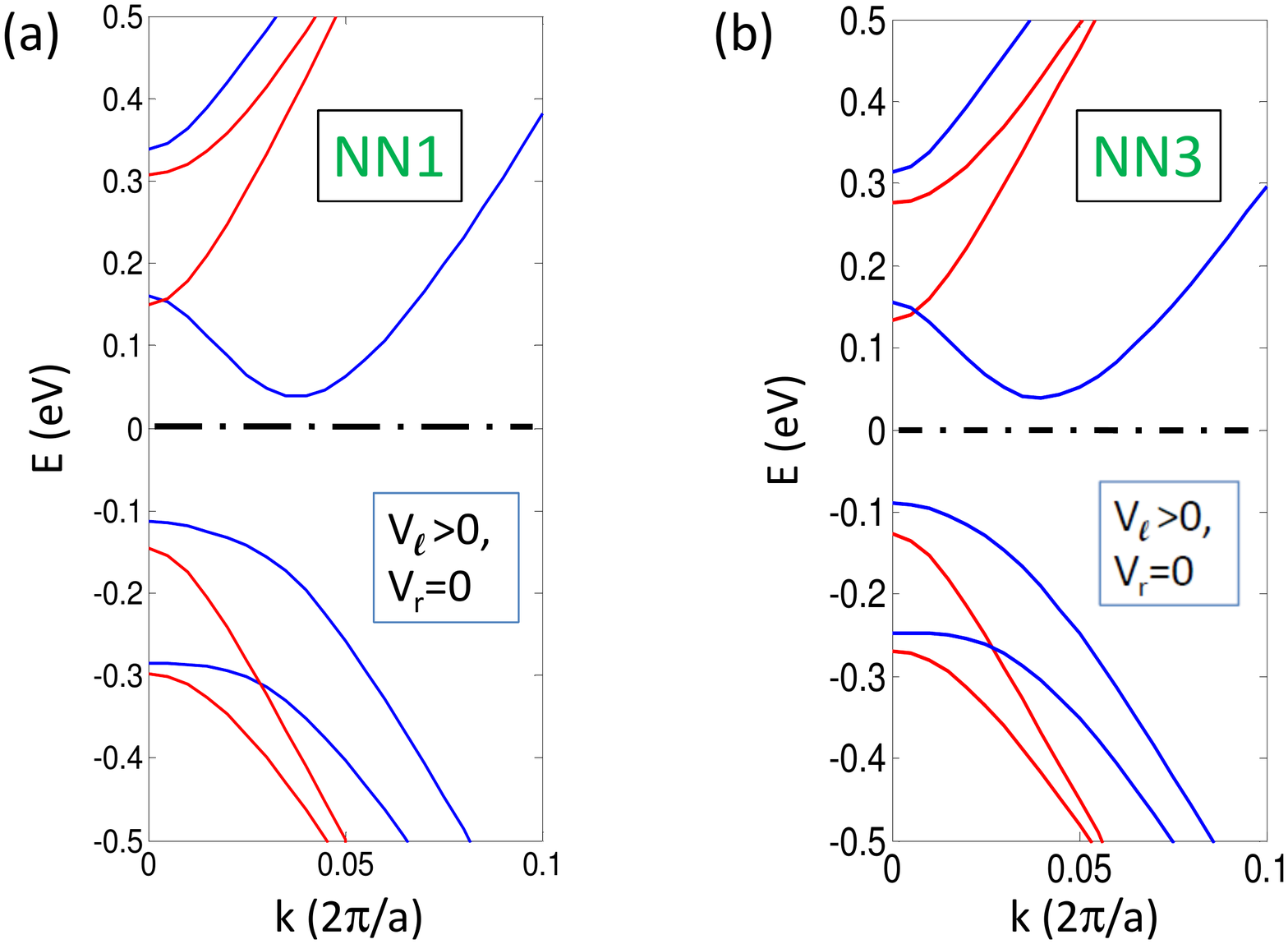}
\caption{\textbf{Supporting Information:} The tight binding
bandstructure of the semiconducting A-GNR with $N=36$ under external
bias conditions, $V_l$=$V_g$=2.8V, $V_r$=0V considering (a): nearest
neighbor interaction, and (b): three nearest neighbor interactions.
The $E$-$k$ relationships (in blue) shows that the conduction band
minimum is shifted from $k$=0, while the valence band maximum
remains at $k$=0. The red lines show the bandstructure without any
external bias. The Chemical potential is set to zero. The nearest
neighbor calculation is found to accurately predict the direct to
indirect bandgap transition.}\nonumber \ec
\end{figure}

\begin{thebibliography}{10}
\bibitem{ksn04}
Novoselov, K. S.; et al. Electric Field Effect in Atomically Thin
Carbon Films. Science \textbf{2004}, 306, 666-669.
\bibitem{ksn05}
Novoselov, K. S.; et al. Two-dimensional gas of massless Dirac
fermions in graphene. Nature \textbf{2005}, 438, 197.
\bibitem{akg07}
Geim, A. K.; et al. The Rise of Graphene. Nat. Mat. \textbf{2007},
6, 183.
\bibitem{ksn06}
Novoselov, K. S.; et al. Unconventional quantum Hall effect and
Berry's phase of $2\pi$ in bilayer graphene. Nat. Phys.
\textbf{2006}, 2, 177.
\bibitem{mcl07}
Lemme, M. C.; et al. A Graphene Field-Effect Device. IEEE Elec. Dev.
Lett. \textbf{2007}, 28, 282.
\bibitem{jbs08}
Oostinga, J. B.; et al. Gate-induced insulating state in bilayer
graphene devices. Nat. Mat. \textbf{2008}, 7, 151.
\bibitem{jhc08}
Chen, J. H.; et al. Intrinsic and extrinsic performance limits of
graphene devices on SiO$_2$. Nat. Nano. \textbf{2008}, 3, 206.
\bibitem{yml10}
Lin, Y. M.; et al. 100-GHz Transistors from Wafer-Scale Epitaxial
Graphene. Science \textbf{2010}, 327, 662.
\bibitem{kn96}
Nakada, K.; et al. Edge state in graphene ribbons: Nanometer size
effect and edge shape dependence. Phys. Rev. B \textbf{1996}, 54,
17954.
\bibitem{yws06}
Son, Y. W.; Energy Gaps in Graphene Nanoribbons. Phys. Rev. Lett.
\textbf{2006}, 97, 216803.
\bibitem{lb06}
Brey, L.; et al. Edge states and the quantized Hall effect in
graphene. Phys. Rev. B \textbf{2006}, 73, 195408.
\bibitem{lb106}
Brey, L.; et al. Electronic states of graphene nanoribbons studied
with the Dirac equation. Phys. Rev. B \textbf{2006}, 73, 235411.
\bibitem{bo06}
Obradovic, B.; et al. Analysis of graphene nanoribbons as a channel
material for field-effect transistors. Appl. Phys Lett.
\textbf{2006}, 88, 142102.
\bibitem{qy07}
Yan, Q.; et al. Intrinsic Current-Voltage Characteristics of
Graphene Nanoribbon Transistors and Effect of Edge Doping. Nano
Lett. \textbf{2007}, 7, 1469.
\bibitem{fmr08}
Rojas, F. M.; et al. Performance limits of graphene-ribbon
field-effect transistors. Phys. Rev. B \textbf{2008}, 77, 045301.
\bibitem{son06}
Son, Y. W.; et al. Half-Metallic Graphene Nanoribbons. Nature
\textbf{2006} 444, 347.
\bibitem{dsn07}
Novikov, D. S.; Transverse Field Effect in Graphene Ribbons. Phys.
Rev. Lett. \textbf{2007}, 99, 056802.
\bibitem{hr08}
Raza, H.; et al. Armchair graphene nanoribbons: Electronic structure
and electric-field modulation. Phys. Rev. B \textbf{2008}, 77,
245434.
\bibitem{rs98}
Saito, R.; et al. ``Physical Properties of Carbon Nanotubes," World
Scientific Publishing, \textbf{1998}.
\bibitem{jcs54}
Slater, J.C.; et al. Simplified LCAO Method for the Peroidic
Potential Problem. Phys. Rev. \textbf{1954}, 94, 1498.
\bibitem{sr02}
Reich, S.; et al. Tight-binding description of graphene. Phys. Rev.
B \textbf{2002}, 66, 035412.
\bibitem{mm54}
Magnus, W.; On the exponential solution of differential equations
for a linear operator. Comm. Pure and Appl. Math. \textbf{1954},
VII: 649.
\bibitem{sb09}
Blanes, S.; et. al. Magnus and Fer expansions for matrix
differential equations: The convergence problem. Phys. Rep.,
\textbf{2009}, 470, 151.
\end{thebibliography}
\end{document}